\documentclass[a4paper,reprint]{achemso}
\usepackage{graphicx}
\usepackage{amsmath}
\usepackage{fancyhdr}
\usepackage{amssymb}
\usepackage{natbib}
\usepackage{cancel}
\usepackage{color}
\setcitestyle{round}

\title{Caliber Corrected Markov Modeling (C$_2$M$_2$): Correcting Equilibrium Markov Models}
\author{Purushottam D. Dixit}
\affiliation{Department of Systems Biology, Columbia University, New York, NY 10032}
\email{dixitpd@gmail.com}

\author{Ken A. Dill}

\affiliation{Laufer Center for Quantitative Biology, Department of Chemistry, and Department of Physics and Astronomy, Stony Brook University, Stony Brook, NY, 11790}

\begin{document}

\begin{abstract}
Rate processes are often modeled using Markov-State Models (MSM).  Suppose you know a \emph{prior} MSM, and then learn that your prediction of some particular observable rate is wrong.  What is the best way to correct the whole MSM?  For example, molecular dynamics simulations of protein folding may sample many microstates, possibly giving correct pathways through them, while also giving the wrong overall folding rate, when compared to experiment.  Here, we describe {\bf C}aliber {\bf C}orrected {\bf M}arkov {\bf M}odeling (C$_2$M$_2$): an approach based on the principle of maximum entropy for updating a Markov model by imposing state- and trajectory-based constraints.  We  show that such corrections are equivalent to asserting position-dependent diffusion coefficients in continuous-time continuous-space Markov processes modeled by a Smoluchowski equation.  We derive the functional form of the diffusion coefficient explicitly in terms of the trajectory-based constraints.  We illustrate with examples of 2D particle diffusion and an overdamped harmonic oscillator.
\end{abstract}
\maketitle

\section{The problem:  Correcting Markov Models from data}

Consider the following type of problem.  You have a network of states $a = 1, 2, \ldots$.  You have a Markov model with known \emph{a priori} transition probabilities $P_{ab}$ between all pairs of states.  Now, you learn from data that some single global average rate quantity predicted by this model is incorrect.  What is the `best' way to correct the full transition matrix to bring it into consistency with the new limited information?  This is a common problem.  First, Markov models are ubiquitous.  Among many other things, they are used to study folding, binding and mechanisms of action of biomolecules~\citep{chodera2014markov}, chemical and biochemical reaction networks~\citep{gillespie,paulsson2004summing}, and the evolutionary dynamics of organisms~\citep{dixit2017recombination}.  Second, such models often require many states, and yet are faced with limited experimental data, or limited physical insights that can constrain the model.

Here's an example.  Computer simulations of proteins identify several different metastable conformational states.  The simulated dynamics among these states can then be captured in Markov State Models (MSMs)~\citep{chodera2014markov}.  But, the underlying forcefields are imperfect, so global rate quantities found from molecular dynamics simulations -- such as their folding times, or  rates along  dominant reaction coordinates -- are often found to be in error.  Because the MSMs are likely to be mimicking the relative rates of microscopic processes, correcting the MSMs to agree with one or more experimental observables is likely to approximate well the full microscopic kinetics.  Here, we describe a solution to this problem, `Caliber Corrected Markov Modeling' (C$_2$M$_2$), which employs the principle of Maximum Caliber~\citep{Press2012}, the dynamical version of the Maximum Entropy principle of inference.

 Prior work solves a related problem of correcting an equilibrium distribution.  Pitera and Chodera~\citep{pitera2012use} developed an approach to fix equilibrium distributions based on experimental constraints.  However, we need a different approach here, for two reasons: (1) There are infinitely many Markov models of the dynamics that are consistent with a given equilibrium distribution.  And: (2) our focus is on the dynamics, because we are interested in biological mechanisms, not just the equilibrium states.  For example, a drug's efficacy is often determined by the dynamics of its unbinding from target proteins, not just its equilibrium binding strength~\citep{copeland2016drug,tiwary2017and}.

Recently, we developed a computational framework to update out of equilibrium Markov models using the maximum relative path entropy (minimum Kullback-Leibler divergence)~\citep{dixit2017out}. We `updated' a 'prior' Markov model such that the updated model was consistent with imposed constraints and was {\it minimally deformed} with respect to the prior  model. We showed that imposition of state- and dynamical trajectory-based constraints changes both the stationary distribution as well as the transition probabilities of the Markov model. However, there is a crucial difference between out of equilibrium processes and equilibrium processes. At equilibrium, the entire stationary distribution is known independently of the dynamics (for example, the Boltzmann distribution) and provides additional constraints for entropy maximization.  Moreover, at equilibrium, the distribution satisfies detailed balanced with respect to the transition probabilities.

In the present work, we address the following general question: how do we update a detailed-balanced equilibrium Markov model so that it satisfies with user-imposed equilibrium and dynamical constraints?  Specifically, we seek a Caliber Corrected Markov Model (C$_2$M$_2$) that reproduces the imposed constraints {\it and} has the maximum relative path entropy (or a minimum Kullback-Leibler divergence) with respect to a prior Markov model.

We first review the work of Pitera and Chodera~\citep{pitera2012use} which serves as the first step in our theoretical development. Consider a system with discrete states $\{a, b, c, \dots \}$ with Hamiltonian $H$ at thermal equilibrium with its surroundings.  The equilibrium distribution over states is given by $x_a \propto e^{-\beta H(a)}$. Here $\beta$ is the inverse tempearture.  Conisder a state-dependent property $f(a)$, for example, the end-to end distance of a peptide. The ensemble average $\langle f \rangle_x$ is given by
\begin{eqnarray}
        \langle f \rangle_x = \sum_a x_a f(a). 
\end{eqnarray}

Imagine a situation where the model prediction $\langle f \rangle_x$ does not agree with the corresponding experimentally measured ensemble average $\bar f$. How do we then update the equilibrium distribution $x_a \rightarrow y_a$ (or equivalently the Hamiltonian $H$) such that the biased distribution $\{ y_a \}$ reproduces $\bar f$? Pitera and Chodera~\citep{pitera2012use} seeked an updated equilibrium distribution $\{ y_a \}$ that least deviated from the prior distribution  $\{ x_a\}$ while reproducing the ensemble average $\langle f \rangle_y = \bar f$. They invoked the principle of  maximum relative entropy (minimum Kullback-Leibler divergence). Briefly, one maximizes the relative entropy
\begin{eqnarray}
        S = -\sum_a y_a \log \frac{y_a}{x_a} \label{eq:ent_st}
\end{eqnarray}
subject to constraint
\begin{eqnarray}
\sum_a y_a f(a) = \bar f
\end{eqnarray}
and $\sum_a y_a = 1$. Carrying out the maximization using Lagrange multipliers~\citep{pitera2012use},
\begin{eqnarray}
y_a \propto x_a e^{-\lambda f(a)} \propto e^{-\beta H(a) - \lambda f(a)}.\label{eq:me0}
\end{eqnarray}
The Lagrange multiplier $\lambda$ dictates the deviation in the prediction of $\langle f \rangle$ between the unbiased ensemble $\{ x_a \}$ and the biased ensemble $\{ y_a \}$. For example, if $\lambda = 0$, $y_a = x_a$.

How do we impose similar biases in dynamics? Below, we develop our maximum entropy framework by updating a continuous time continuous space Smoluchowski equation for a particle diffusion on a one dimensional free energy landscape. Generalizations to continuous time discrete space and discrete time discrete space models are presented along the way.

\section{Example of a particle diffusing with bias along one dimension}
\begin{figure}
        \includegraphics[scale=0.4]{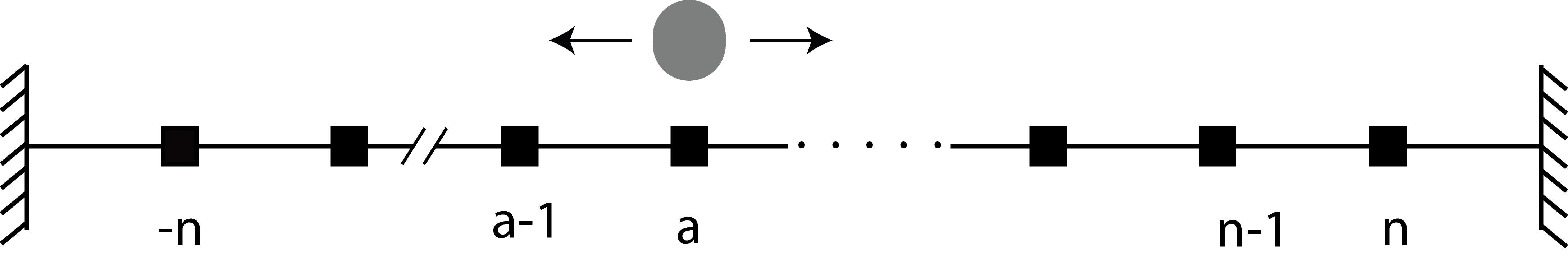}
        \caption{{\bf Diffusion of a particle in one dimension}. Discretized scheme of a particle diffusing in one dimension between $a=-L$ and $a = L$ divi
ded in $2n+1$ nodes labeled from $-n$ to $+n$. \label{fg:1d}}
\end{figure}

Consider a particle diffusing in one dimension between $a \in [-L, L]$ (see Fig.~\ref{fg:1d}). If $p(a,t)$ is the instantaneous probability distribution and
 $x_a \propto \exp \left ( -\beta F(a) \right )$ is the equilibrium distribution, the dynamics of $p(a,t)$ are described by the diffusion equation:\begin{eqnarray}\frac{\partial}{\partial t}p(a,t) = D_0\frac{\partial}{\partial a} \left \{ e^{ -\beta F(a) } \frac{\partial}{\partial a} \left [e^{ \beta F(a) } p(a,t) \right ]\right \} \nonumber  \label{eq:smol1}\\
\end{eqnarray}

In Eq.~\ref{eq:smol1}, $D_0$ is the diffusion coefficient and $F(a)$ is the free energy surface.  The diffusion coefficient sets the time scale of the system. In practice, it can be determined as the constant of proportionality between the ensemble average of the variance of the displacement $a(t) - a(0)$ and the time $t$~\citep{woolf1994conformational,hummer2005position}.  We can discretize the partial differential equation in steps of $da$ and write a continuous-time discrete-state Markov process~\citep{bicout1998electron,hummer2005position},
\begin{eqnarray}
\frac{d}{dt} p(a,t) &=& \omega_{a-da,a} p(a-da,t) + \omega_{a+da,a} p(a+da,t) \nonumber \\ &-& p(a,t)\left (\omega_{a,a+da} + \omega_{a,a-da}  \right ) \label{eq:master}
\end{eqnarray}
where the {\it transition rates} are given by~\citep{bicout1998electron,hummer2005position}
\begin{eqnarray}
\omega_{a,a+da} = D_0\sqrt{\frac{x_{a+da}}{x_a}}  \left (da^2 \right ). \label{eq:omega_sm}
\end{eqnarray}

In Eq.~\ref{eq:smol1}, $D_0$ is the diffusion coefficient and $F(a)$ is the free energy surface.  The diffusion coefficient sets the time scale of the system. In practice, it can be determined as the constant of proportionality between the ensemble average of the variance of the displacement $a(t) - a(0)$ and the time $t$~\citep{woolf1994conformational,hummer2005position}.  We can discretize the partial differential equation in steps of $da$ and write a continuous-time discrete-state Markov process~\citep{bicout1998electron,hummer2005position},
\begin{eqnarray}
\frac{d}{dt} p(a,t) &=& \omega_{a-da,a} p(a-da,t) + \omega_{a+da,a} p(a+da,t) \nonumber \\ &-& p(a,t)\left (\omega_{a,a+da} + \omega_{a,a-da}  \right ) \label{eq:master}
\end{eqnarray}
where the {\it transition rates} are given by~\citep{bicout1998electron,hummer2005position}
\begin{eqnarray}
\omega_{a,a+da} = D_0\sqrt{\frac{x_{a+da}}{x_a}}  \left (da^2 \right ). \label{eq:omega_sm}
\end{eqnarray}

Eq.~\ref{eq:master} can be time-discretized using a small time interval $dt$. We write
\begin{eqnarray}
p(a,t+dt) = \sum_{b} P_{ba} \times  p(b,t)
\end{eqnarray}
where  the {\it transition probabilities} are given by
\begin{eqnarray}
P_{ab} = \omega_{ab} dt~{\it if~}b\neq a \label{eq:pab_approx}
\end{eqnarray}
and
\begin{eqnarray}
P_{aa} = 1- \sum_{b\neq a}  P_{ab} \label{eq:pab_norm}
\end{eqnarray}
Eq.~\ref{eq:pab_norm} ensures that probabilities are conserved and normalized throughout the time evolution.

We carry out the desired state- and trajectory-based biasing for the discrete time discrete state Markov model and then take appropriate continuous limits. First, we comment on the nature of trajectory-based observables. Consider a dynamical variable $r_{ab}$ that is defined over individual transitions of a trajectory of the Markov process. An example of $r_{ab}$ is the number of contacts formed/broken in a single time step by a polymer.  The average $r_{ab}$ over an ensemble of stationary state trajectories is given by~\citep{dixit2014inferring,dixit2015inferring,dixit2015stationary,dixit2017out}
\begin{eqnarray}
\langle r \rangle = \sum_{a,b} x_a P_{ab} r_{ab}. \label{eq:rab_0}
\end{eqnarray}

We want to modify the Markov model described by Eq.~\ref{eq:pab_approx} such that the updated Markov model (with transition probabilities $\{ k_{ab} \}$)  has its equilibrium distribution equal to $\{ y_a \}$ and reproduces the trajectory-ensemble average $\bar r$ which is different than $\langle r_{ab} \rangle$ defined in Eq.~\ref{eq:rab_0}. Previously, Wan et al.~\citep{wan2016maximum} have addressed the problem of updating Markov processes by updating their equilibrium distribution alone (see also Zhou et al.~\citep{zhou2017}).  We proceed by maximizing the relative entropy~\citep{rached2004kullback,wan2016maximum,dixit2017out}
\begin{eqnarray}
S = -\sum_{a,b} y_a k_{ab} \log \frac{k_{ab}}{P_{ab}}
\end{eqnarray}
subject to constraint
\begin{eqnarray}
\sum_a y_a k_{ab}r_{ab} = \bar r \label{eq:dyna}
\end{eqnarray}
and
\begin{eqnarray}
\sum_{b} y_a k_{ab} = y_a~\forall~a,\label{eq:constr1}\\
\sum_{a} y_a k_{ab} = y_b~\forall~b,{\rm ~and}\label{eq:constr2}\\
y_a k_{ab} = y_b k_{ba}~\forall~a~{\rm and}~b.\label{eq:constr3}
\end{eqnarray}
Eq.~\ref{eq:constr1} ensures that state probabilities are conserved and normalized throughout the time evolution of the Markov process. Eq.~\ref{eq:constr2} imposes $\{ y_a \}$ as the stationary distribution of the Markov process. Finally, Eq.~\ref{eq:constr3} explicitly imposes microscopic detailed balance.

We write the Caliber $C$ by incorporating these constraints using Lagrange multipliers~\citep{Press2012,dixit2014inferring,dixit2015inferring,dixit2015stationary,wan2016maximum}
\begin{eqnarray}
C &=&  -\sum_{a,b} y_a k_{ab} \log k_{ab} +  \sum_{a,b} y_a k_{ab} \log P_{ab} \nonumber \\&+& \sum_a l_a \left ( \sum_b y_a k_{ab} - y_a\right ) +  \sum_b m_b \left ( \sum_a y_a k_{ab} - y_b\right ) \nonumber \\
&+& \sum_{a,b} \epsilon_{ab}\left (y_a k_{ab} - y_{b}k_{ba} \right ) + \gamma \left (  \sum_{a,b} y_{a} k_{ab} r_{ab} - \bar r\right ). \label{eq:caliber}
\end{eqnarray}
In Eq.~\ref{eq:caliber},  Lagrange multipliers $\{ l_a \}$ impose constraints in Eq.~\ref{eq:constr1}.  $\{ m_b \}$ impose constraints in Eq.~\ref{eq:constr2}. $\{ \epsilon_{ab} \}$ impose microscopic detailed balance. Finally,  $\gamma$ imposes the constraint of ensemble average $\bar r$ of the dynamical variable $r_{ab}$ (see Eq.~\ref{eq:dyna}).

Differentiating with respect to $k_{ab}$ and setting the derivative to zero and imposing the detailed balance constraint we have (see appendix~\ref{ap1})
\begin{eqnarray}
k_{ab} = \frac{\mu_a \mu_b}{y_a}G_{ab} \label{eq:k1}
\end{eqnarray}
where
\begin{eqnarray}
G_{ab} = \frac{P_{ab}}{x_b}\exp \left (\gamma \frac{r_{ab} + r_{ba}}{2}\right).\label{eq:G}
\end{eqnarray}
For a specific value of the Lagrange multiplier $\gamma$, we determine the modified Lagrange multipliers $\mu_a$ by imposing the constraints given in Eq.~\ref{eq:constr1}. We have~\citep{dixit2014inferring,dixit2015inferring}
\begin{eqnarray}
\sum_b k_{ab} &=& 1~\forall~a\\
\Rightarrow   \sum_b G_{ab} \mu_b &=&\frac{y_a}{\mu_a}~\forall~a. \label{eq:interm}
\end{eqnarray}
Eq.~\ref{eq:interm} can be reorganized by defining a non-linear operator $D[\bar \mu]_a = y_a/\mu_a$. We note that $D[ D[\bar \mu]] = \bar \mu$. We have
\begin{eqnarray}
G\bar \mu = D[\bar \mu] \Rightarrow D[G\bar \mu] = \bar \mu.\label{eq:fixd}
\end{eqnarray}
In other words, the vector $\bar \mu$ of modified Lagrange multipliers can be numerically solved as a fixed point of Eq.~\ref{eq:fixd}. Note that the matrix $G$ is symmetric since the transition probabilities $\{ P_{ab} \}$ satisfy detailed balance with respect to the equilibrium distribution$\{ x_a \}$. Eq.~\ref{eq:G} indicates that an important consequence of imposing detailed balance~\cite{dixit2014inferring,dixit2015inferring,dixit2015stationary} is that the dynamical constraint $r_{ab}$ appears in its symmetrized form $(r_{ab} + r_{ba})/2$. From now onwards, for simplicity, we assume the constraint $r_{ab}$ is already symmetrized in $a$ and $b$; $r_{ab} = r_{ba}$ for all $a$ and $b$.

In the limit $dt\rightarrow 0$ in Eq.~\ref{eq:pab_approx}, the modified Lagrange multipliers $\{\mu_a \}$ can be solved analytically. In this limit, the transition probabilities $k_{ab}$ of the updated Markov process are given by (see appendix~\ref{ap2})
\begin{eqnarray}
k_{ab} = dt\sqrt{\frac{y_b}{y_a}}\sqrt{\frac{x_a}{x_b}}\omega_{ab} \exp \left ( \gamma r^{\dag}_{ab} \right )~{\it if}~a\neq b \label{eq:kab_dt}
\end{eqnarray}
where (see appendix~\ref{ap2} for details)
\begin{eqnarray}
r^{\dag}_{ab} =\frac{r_{ab} + r_{ba}}{2} - \frac{r_{aa}+r_{bb}}{2}.\label{eq:rdag}
\end{eqnarray}
Eq.~\ref{eq:G} and Eq.~\ref{eq:rdag} indicate that when we impose detailed balance and take the continuous time limit, we modify the dynamical constraint  to make it symmetric in $a$ and $b$ and to have $r_{aa} = 0~\forall~a$. Mathematically, we perform the  transformation given by Eq.~\ref{eq:rdag}. For brevity, from now onwards, we assume that this transformation has already been performed (unless specified otherwise). We drop the $\dag$ superscript for simplicity.

Before we proceed further, let us examine the consequences of the transformation in Eq.~\ref{eq:rdag}. Consider an antisymmetric  dynamical quantity  such that $r_{ab} + r_{ba} = 0$. A constraint which imposes a finite value of $\langle r_{ab} \rangle$ is clearly inconsistent with detailed balance. Indeed, the transformation in Eq.~\ref{eq:rdag} modifies $r_{ab}$ to $r^{\dag}_{ab} = 0$. Similarly,  consider when $r_{ab} = \lambda_1 f(a) + \lambda_2 g(b)$ (where $\lambda_1$ and $\lambda_2$ are constants and $f$ and $g$ are state-dependent functions) can be separated as a sum of two state-based constraints. Since we explicitly constrain the stationary distribution $\{ y_a\}$, we do not have additional freedom to constrain state-dependent quantities. Here too, the transformation modifies the constraint to $r^{\dag}_{ab} = 0$.

From Eq.~\ref{eq:kab_dt}, the updated continuous time {\it transition rates} $\kappa_{ab}$ are given by
\begin{eqnarray}
\kappa_{ab} = \lim_{dt\rightarrow 0} \frac{k_{ab}}{dt} = \sqrt{\frac{y_b}{y_a}}\sqrt{\frac{x_a}{x_b}}\omega_{ab} \exp \left ( \gamma r_{ab} \right ) \label{eq:kappa}
\end{eqnarray}
In Eq.~\ref{eq:kappa}, the {\it transition rates} $\kappa_{ab}$ describe an updated Markov process that is minimally biased with respect to the prior Markov process given by rates $\omega_{ab}$ (see Eq.~\ref{eq:omega_sm}) and a) has a prescribed equilibrium distribution $\{ y_a \}$ and b) reproduces a prescribed dynamical average $\langle r_{ab} \rangle$.

Next we  take the continuous space limit of Eq.~\ref{eq:kappa} by substituting $\omega_{ab}$ given in Eq.~\ref{eq:omega_sm}. We have
\begin{eqnarray}
\kappa_{a,a+da} = \sqrt{\frac{y_{a+da}}{y_a}}  D({a+da/2}) \left (da^2 \right ) \nonumber \label{eq:kappa_sm} \\
\end{eqnarray}
where
\begin{eqnarray}
D({a+da/2})  = D_0 \exp \left ( \gamma r_{a,a+da} \right ) \label{eq:modD}
\end{eqnarray}
is the updated diffusion coefficient at  $a + da/2$ and $y_a\propto e^{-\beta G(a)}$ is the prescribed equilibrium distribution. The exponential $\exp \left ( \gamma r_{a,a+da} \right )$ will have a non-trivial contribution to the diffusion coefficient {\it only if} $\gamma \propto 1/da^n$. We assume that $r_{a,a+da} = h(a)da^n$ and $\gamma = \gamma_0 /da^n$. Here, $\gamma_0 = o(1)$.  We have
\begin{eqnarray}
D({a+da/2})  = D_0 \exp \left ( \gamma_0 h(a)\right ). \label{eq:D_define}
\end{eqnarray}

Below, we show how to explicitly derive $D(a)$ from the functional form of the constraints.
Comparing Eq.~\ref{eq:kappa_sm} and Eq.~\ref{eq:omega_sm}, the biased Smoluchowski equation is given by\begin{eqnarray}\frac{\partial}{\partial t}p(a,t) = \frac{\partial}{\partial a} \left \{  D(a) e^{ -\beta G(a) } \frac{\partial}{\partial a} \left [e^{ \beta G(a) } p(a,t) 
\right ]\right \} \nonumber  \label{eq:smol2}\\
\end{eqnarray}
where
\begin{eqnarray}
D(a) = D_0e^{\gamma h(a)} \label{eq:d_pos}\end{eqnarray}is the position-dependent diffusion coefficient.

Before we further illustrate Eq.~\ref{eq:smol2} with examples, we make a few observations. First, if we only update the equilibrium distribution ($x_a \rightarrow y_a$) and impose no additional dynamical constraint, the corresponding change in the Smoluchowski equation is simply changing its equilibrium distribution (see Eq.~\ref{eq:smol1} and Eq.~\ref{eq:smol2}). In contrast, the imposition of trajectory-based constraints leads to a diffusion coefficient $D(a)$ that depends on the position $a$. Second, a straightforward modification allows us to incorporate multiple dynamical constraints. Each constraint is associated with one Lagrange multiplier. In this case, the diffusion coefficient is given by
\begin{eqnarray}
D(a) = D_0e^{\sum_i \gamma_i h_i(a)} \label{eq:d_pos1}
\end{eqnarray}

As an illustration for the recipe to calculate diffusion coefficients from trajectory-based constraints, let us also look at specific constraints. Consider $r_{ab} = \phi(a)\phi(b)$ for some function $\phi$ of the position. The constraint $\langle r_{ab} \rangle$ represents the autocorrelation of the quantity $\phi$ along dynamical trajectories of the Markov process.  After performing the transformation in Eq.~\ref{eq:rdag}, we have (omiting the $\dag$ for brevity)
\begin{eqnarray}
r_{ab} = -(\phi(a)-\phi(b))^2/2. \label{eq:rmod}
\end{eqnarray} Thus from Eq.~\ref{eq:d_pos}
\begin{eqnarray}
r_{a,a+da} &=&-1/2\phi'(a)da^2 \nonumber \Rightarrow D(a) = D_0 e^{-\gamma \phi'(a)^2/2} \nonumber \\&\approx& \frac{D_0}{1+ \gamma \phi'(a)^2/2}. \label{eq:zw1}
\end{eqnarray}
The second approximation holds true when $|\gamma| \ll1$ or when $\phi(a)$ is slowly varying.

Two notable examples of constraints are (1) a position-position autocorrelation function, and (2) a PMF-PMF autocorrelation along a stochastic trajectory. These constraints can be represented as $\langle a b \rangle$ and $\langle F(a)F(b) \rangle$ respectively. Here, $a$ is the position coordinate and $F(a)$ is the corresponding free energy. When we constrain the position-position autocorrelation, the updated diffusion coefficient does not depend on the position but simply takes a different value than the `prior' diffusion coefficient. We have
\begin{eqnarray}
D(a) = D_0 e^{-\gamma}.\label{eq:dpos}
\end{eqnarray}
 In contrast, when we constrain the PMF-PMF correlation, we have
\begin{eqnarray}
D(a) = D_0 e^{-\gamma F'(a)^2/2} &\approx& \frac{D_0}{1 + \gamma F'(a)^2/2} \\ 
&=& \frac{D_0}{1 + \gamma f_a^2/2} \label{eq:dpmf}
\end{eqnarray}
where we have identified $f_a = -F'(a)$ as the average force at position $a$.

Position-dependent diffusion coefficients have been interpreted as effective corrections to lower dimensional projections of higher dimensional dynamics~\citep{zwanzig1992diffusion,berezhkovskii2011time}. Specifically, Zwanzig~\citep{zwanzig1992diffusion} showed that the one dimensional diffusive dynamics along the length of a two dimensional channel with variable width $w(a)$ is best described by a position-dependent diffusion coefficient
\begin{eqnarray}
D(a) = \frac{D_0}{1+w'(a)^2/12}. \label{eq:zw2}
\end{eqnarray}
From Eq.~\ref{eq:zw1} and Eq.~\ref{eq:zw2}, this result can be interpreted within the maximum relative entropy framework as a position dependent diffusion coefficient arising from the dynamical constraint $\langle r_{ab} \rangle$ where $r_{ab} = w(a)w(b)$ (width-width correlation along stochastic trajectories). Berezhkovskii and Szabo~\citep{berezhkovskii2011time} considerably generalized the original work by Zwanzig and explicitly derived the formula for the position dependent diffusion coefficient for diffusion along a `slow' dimension in a multi-dimensional system.

In recent years, position-dependent diffusion coefficients have proved to be a very popular in studying lower-dimensional dynamics of complex biomolecules.  For example, Best and Hummer~\citep{best2006diffusive,best2010coordinate} have studied the effective dynamics of protein folding along a one-dimensional reaction coordinates defined as the fraction of native contacts, Chodera and Pande~\citep{chodera2011splitting} have studied unfolding of a DNA hairpin along the extension of the hairpin. In many such examples, the central goal is to infer the position-dependent diffusion coefficient from molecular dynamics data. Complementary to these studies, in this work we interpret position dependent diffusion coefficient as arising from trajectory-based constraints on Markovian dynamics.

\section{Multidimensional problem: Illustration in two dimensions}

\begin{figure}
        \includegraphics[scale=0.4]{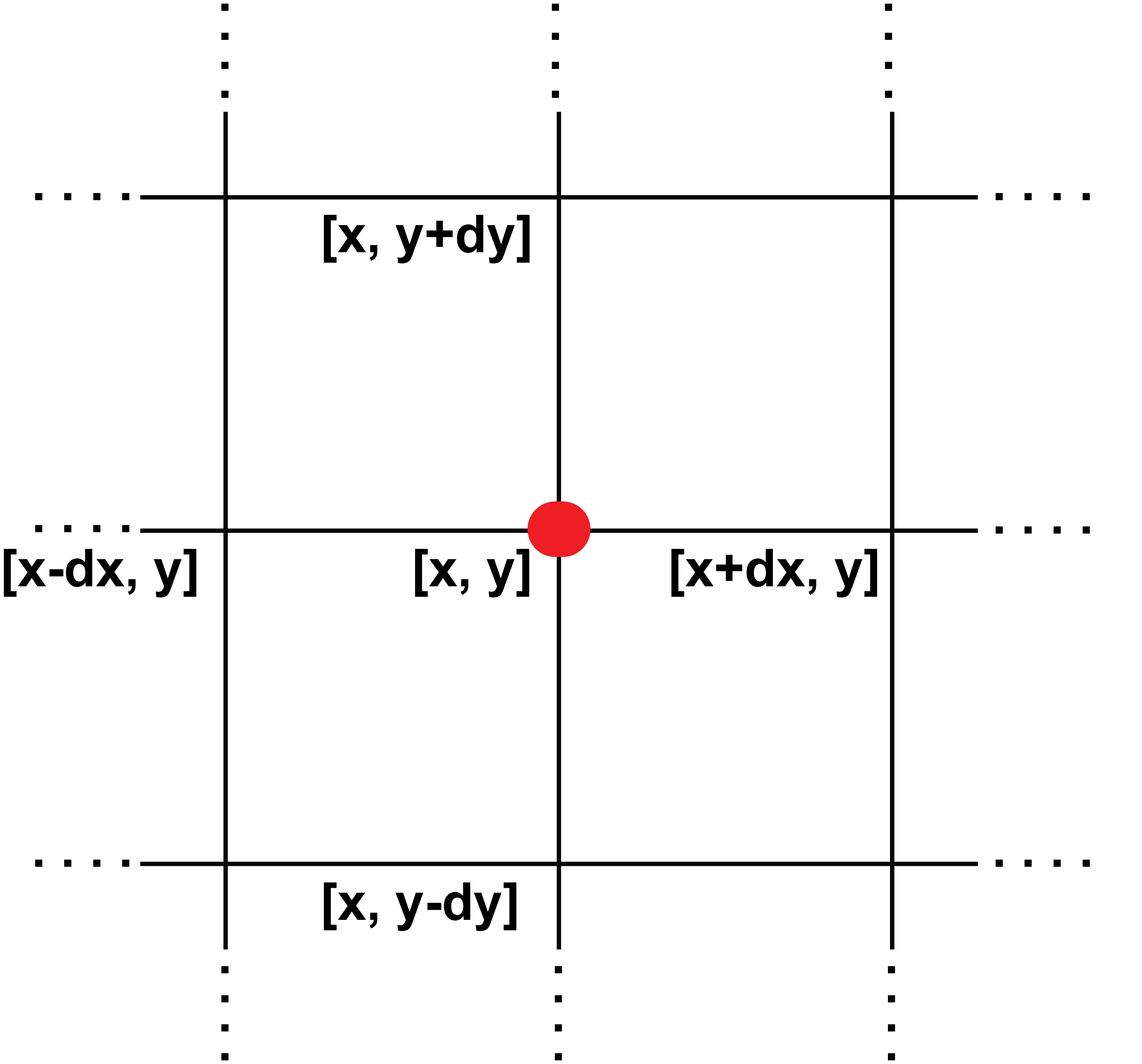}
        \caption{{\bf Discretization of a two dimensional diffusion problem}. A particle diffusing on a two dimensional lattice with spacing $dx = dy$ in $x$ and $y$ directions. In a single time step, the particle can hop to one of its nearest neighbors shown in the figure. \label{fg:2dlattice}}
\end{figure}
In the above development,  we updated the Smoluchowski equation in one dimension. However, the method developed can be generalized to a multidimensional problem in a straightforward manner. We note that the rest of the manuscript can be read without this section.

We illustrate the two dimensional derivation with a particle diffusing on a two dimensional landscape (see Fig.~\ref{fg:2dlattice}. For simplificty of notation, we assume that the energy landscape and equilibrium probability distribution of the `prior' process is flat; $p_{\rm eq}([x, y]) = const$. Let $p([x,y]|t)$ denote the probability of observing the particle at position $[x,y]$ at time $t$. The `prior' dynamics of $p([x,y]|t)$ is given by
\begin{eqnarray}
\frac{\partial}{\partial t} p([x,y]|t) &=& D_x\frac{\partial^2}{\partial x^2}p([x,y]|t) + D_y\frac{\partial^2}{\partial y^2}p([x,y]|t)  \nonumber \label{eq:2d_diff} \\
\end{eqnarray}
In Eq.~\ref{eq:2d_diff} $D_x$ is the diffusion coefficient in the $x$ direction and $D_y$ is the diffusion coefficient in the $y$ direction.

Consider that we update the prior Markov model given by Eq.~\ref{eq:2d_diff} by imposing  a dynamical constraint $\langle r_{ab} \rangle$ as was done for the one dimensional case above. We also impose that the equilibrium distribution remains unchanged; $p_{\rm eq}([x,y]) = const$.  For simplicity, as above, we assume that $r_{ab}$ quantifies correlation in some quantity $\phi$ along dynamical trajectories. Mathematically, $r_{ab} = \phi(a)\phi(b)$.

As we show in the appendix~\ref{ap3}, imposing a dynamical constraint introduces position dependent coefficient in the 2-dimensional problem as well. We have the updated diffusion coefficients $D_x^{\rm new}$ and $D_y^{\rm new}$:
\begin{eqnarray}
D_x^{\rm new}([x,y]) = D_x\exp \left  (-\gamma_0 \frac{\left ( \frac{\partial }{\partial x} \phi([x,y]) \right )^2}{2} \right )\nonumber \\
D_y^{\rm new}([x,y]) = D_y\exp \left  (-\gamma_0 \frac{\left ( \frac{\partial }{\partial y} \phi([x,y]) \right )^2}{2} \right ) \label{eq:diffusion_2d_mod}
\end{eqnarray}
where $\gamma_0$ is the modified Lagrange multiplier.  Notably, updating diffusion coefficients based on dynamical constraints can introduce dynamical anisotropy. The update to the diffusion coefficient in the $x-$ direction is different from the one  in the $y-$direction.

\section{Example application: overdamped oscillator}

\begin{figure}
       \includegraphics[scale=0.45]{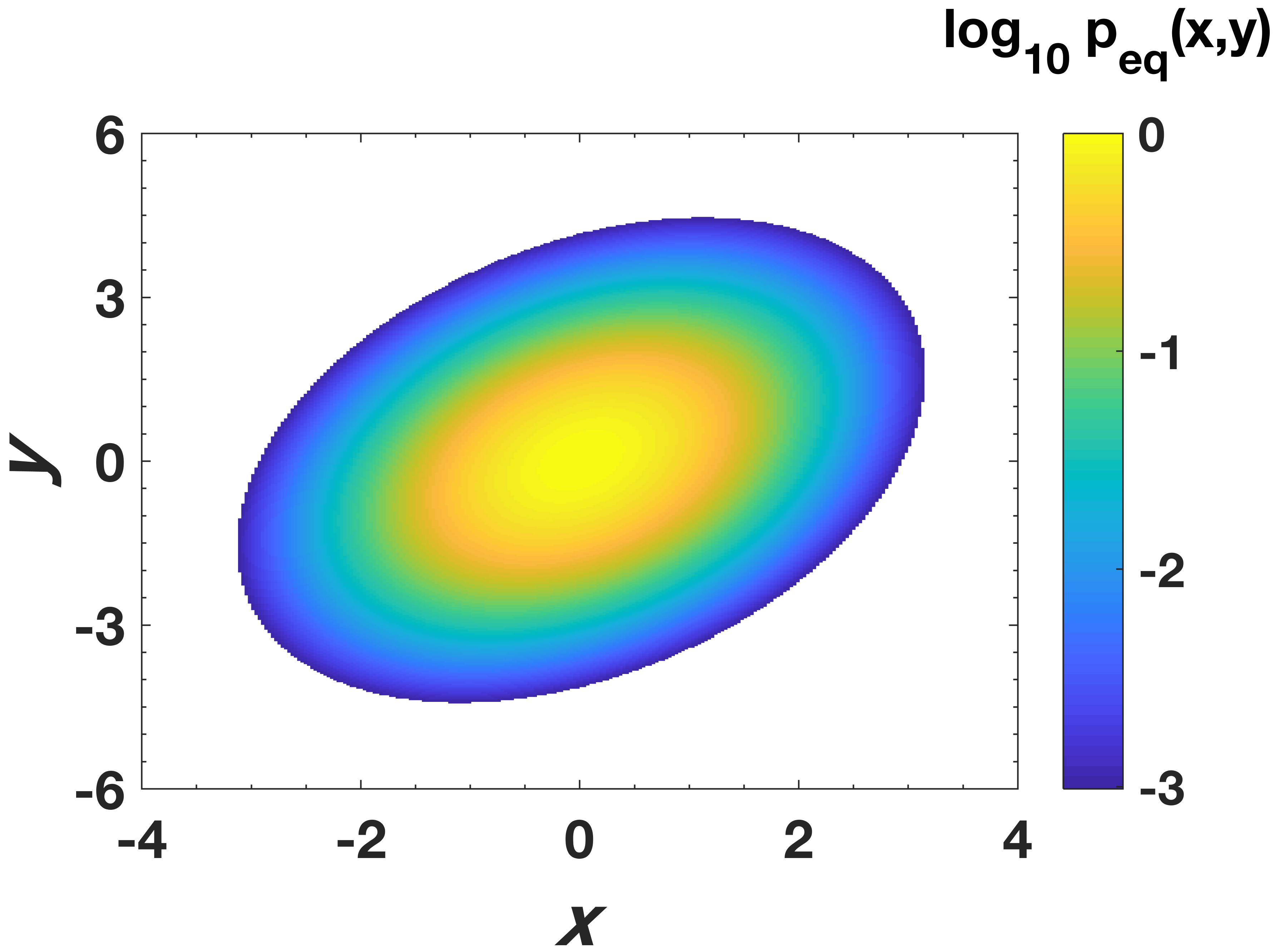}
       \caption{{\bf The equilibrium probability distribution of the overdamped harmonic oscillator.} The equilibrium distribution of the Harmonic oscillator is given by a two dimensional normal distribution (see Eq.~\ref{eq:peq_harm}).\label{fg:peq}}
\end{figure}
We now illustrate an application of present method to an overdamped Harmonic oscillator.  Consider a two-dimensional harmonic oscillator in equilibrium with its thermal surroundings and undergoing overdamped Langevin dynamics. The equations of motion of the 2 dimensional Harmonic oscillator are
\begin{eqnarray}
       \dot x(t) &=& D_x\left (-1.6x(t)  + 0.4y(t)\right ) +\sqrt{2D_x} \eta_1(t) \nonumber \\
       \dot y(t) &=& D_x \left (0.4x(t)  - 0.8y(t)\right ) + \sqrt{2D_y} \eta_2(t) \label{eq:lang}
\end{eqnarray}
where $\langle \eta_i(t)\eta_j(t')\rangle = \Delta_{ij} \delta(t-t')$ for $i,j = 1, 2$. Here, $\Delta_{ij}$ is the Kronecker delta function and $\delta(t)$ is the Dirac delta function. The diffusion constants are $D_x = 5$ and $D_y = 1$ units and the inverse temperature is $\beta = 1$. The equilibrium distribution $p_{\rm eq}(x,y)$ is described by a two dimensional Gaussian distribution,
\begin{eqnarray}
p_{\rm eq}(x,y) \propto e^{-0.8x^2 + 0.4 xy - 0.4 y^2} \label{eq:peq_harm}
\end{eqnarray}
and is shown in Fig.~\ref{fg:peq}. We simulate Eqs.~\ref{eq:lang} with a discretized Langevin dynamics scheme with $dt = 0.001$ units.

\begin{figure}
        \includegraphics[scale=0.4]{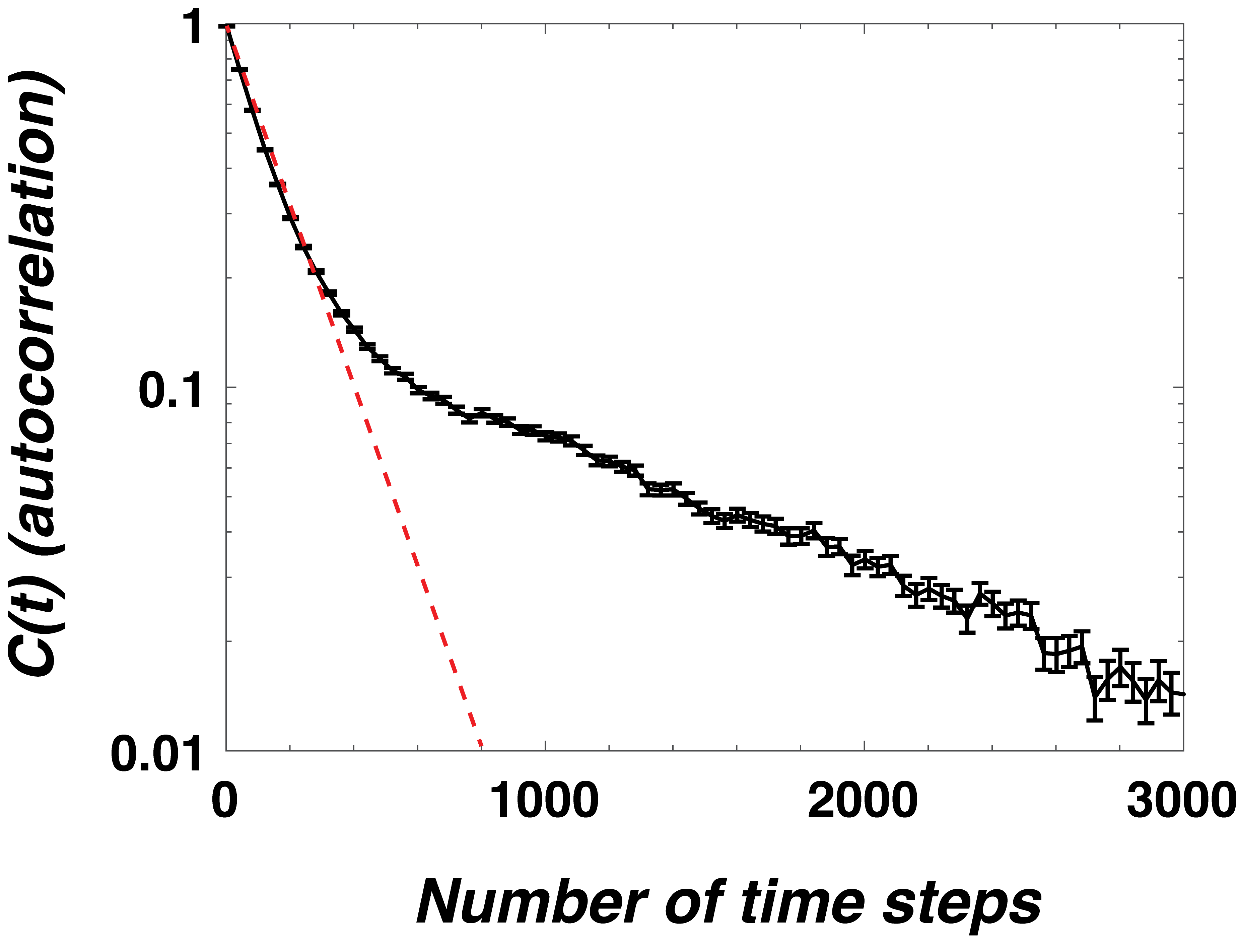}
        \caption{{\bf Autocorrelation function of the two dimensional dynamics reveals two time scales}. The normalized autocorrelation function $C(t)$. The dashed red line indicates the faster of the two time scales in the autocorrelation function. The error bars represent standard deviation in mean estimated 
from 500 independent calculations of $C(t)$. \label{fg:autoc_2d}}
\end{figure}

In Fig.~\ref{fg:autoc_2d} we show the normalized autocorrelation function
\begin{eqnarray}
C(t) = \frac{\langle x(t) x(0) \rangle - \langle x(0) \rangle^2}{\langle x(0)^2 \rangle - \langle x(0) \rangle^2}  \label{eq:autoc}
\end{eqnarray} of the one dimensional projection $x(t)$ of the two dimensional dynamics.  The autocorrelation decays with two time scales, a fast decay for 
$t < 1000dt$ and a slower decay after $t > 1000dt$.

How do we describe the {\it effective} stochastic dynamics of $x(t)$? The marginal equilibrium distribution $p_{\rm eq}(x) = \int p_{\rm eq}(x,y)dy$ is (see Appendix~\ref{ap4})\begin{eqnarray}
p_{\rm eq}(x) \propto e^{-0.7 x^2}.\label{eq:peq_x}
\end{eqnarray}
As a first guess, we write down the simplest Smoluchowski equation that relaxes to this equilibrium distribution. We have
\begin{eqnarray}
\frac{\partial}{\partial t}p(x,t) = D_x\frac{\partial}{\partial x} \left \{ e^{ -0.7x^2 } \frac{\partial}{\partial x} \left [e^{ 0.7x^2 } p(x,t) \right ]\right \}.   \label{eq:smol_x}
\end{eqnarray}
It is well known that the autocorrelation function described by Eq.~\ref{eq:smol_x} decays exponentially with a single time constant~\citep{zwanzig2001nonequilibrium}. As a result Eq.~\ref{eq:smol_x} cannot capture the essential features of $x(t)$ dynamics.

Can we model the dynamics with a position-dependent diffusion coefficient? We impose two kinetic constraints noted above (see Eq.~\ref{eq:dpos} and Eq.~\ref{eq:dpmf}).  The first constraint corresponds to the position-position autocorrelation  and the second constraint corresponds to the PMF-PMF autocorrelation. The corresponding position dependent diffusion coefficient is given by (see Eq.~\ref{eq:zw1})
\begin{eqnarray}
D(x) = D_x \exp \left ( - \gamma_1 -  \gamma_2 x^2 \right ). \label{eq:dx}
\end{eqnarray}
Here, $\gamma_1$ and $\gamma_2$ are Lagrange multipliers that relate to the dynamical constraint $\langle r_{ab} \rangle = \langle ab \rangle$ and $\langle r_{ab} \rangle = \langle F(a) F(b) \rangle$ respectively. As discussed above (see Eq.~\ref{eq:dpos} and Eq.~\ref{eq:dpmf}), the Lagrange multiplier $\gamma_1$ allows us to adjust the overall diffusion constant. The Lagrange multiplier $\gamma_2 > 0$ slows down diffusion in regions of the $x-$space where PMF changes most rapidly. Specifically, diffusion coefficient gets smaller as $|x|$ increases.

The  Smoluchowski equation with a position dependent diffusion coefficient is given by
\begin{eqnarray}
\frac{\partial}{\partial t}p(x,t) = \frac{\partial}{\partial x} \left \{ D(x) e^{ -0.7x^2 } \frac{\partial}{\partial x} \left [e^{ 0.7x^2 } p(x,t) \right ]\right \} \nonumber  \label{eq:smol1} \\
\end{eqnarray}

\begin{figure}
        \includegraphics[scale=0.45]{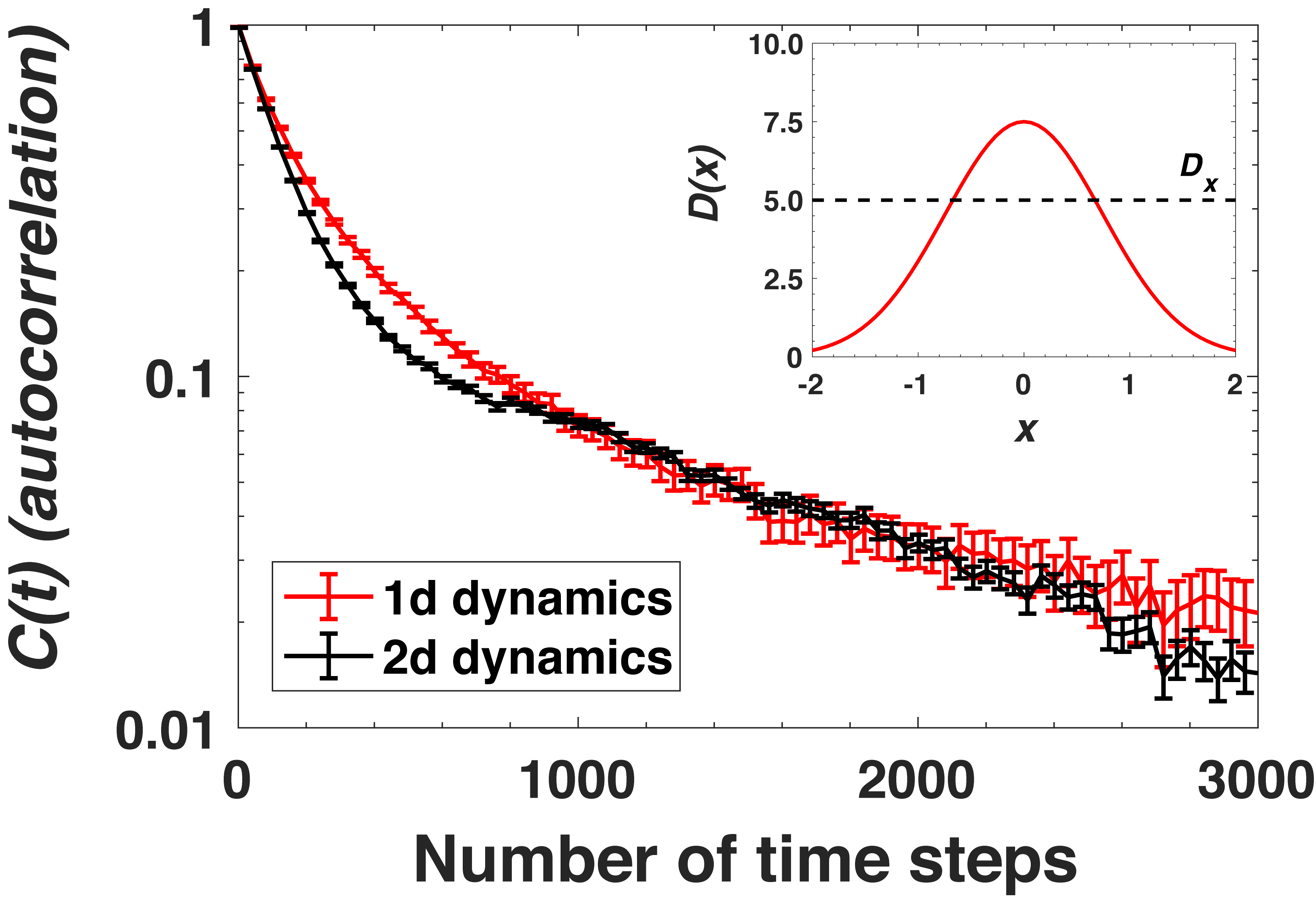}
        \caption{{\bf Effective dynamics with a position dependent diffusion coefficient captures the autocorrelation function of the two dimensional harmonic oscillator}. The comparison between the normalized autocorrelation function $C(t)$ estimated using Eq.~\ref{eq:smol1} (red) and Eq.~\ref{eq:lang} (black). The effective one dimensional dynamics of Eq.~\ref{eq:smol1} captures the two time scales in $x(t)$ dynamics.  The error bars represent standard deviation in mean estimated from 500 independent calculations of $C(t)$. The inset shows the position dependence of the diffusion coefficient $D(x)$ on $x$. \label{fg:autoc_1d}}
\end{figure}
In Fig.~\ref{fg:autoc_1d} we plot the normalized autocorrelation function $C(t)$ (red dots) as predicted by the stochastic dynamics described by Eq.~\ref{eq:smol1} and compare it to the autocorrelation function shown in Fig.~\ref{fg:autoc_2d} (black line).  We have used $\gamma_1 = -0.4$, $\gamma_2 = 0.9$. We have a discretization time step of $dt = 0.001$ and used the Ito convention to simulate the position dependent diffusion coefficient (see appendix~\ref{ap3}). The inset shows the dependence of the diffusion coefficient on $x$. Notably, incorporating a  position dependent diffusion coefficient in Eq.~\ref{eq:dx} allows us to capture the two time scales observed in $x(t)$-dynamics with sufficient accuracy.

\begin{figure}
        \includegraphics[scale=0.45]{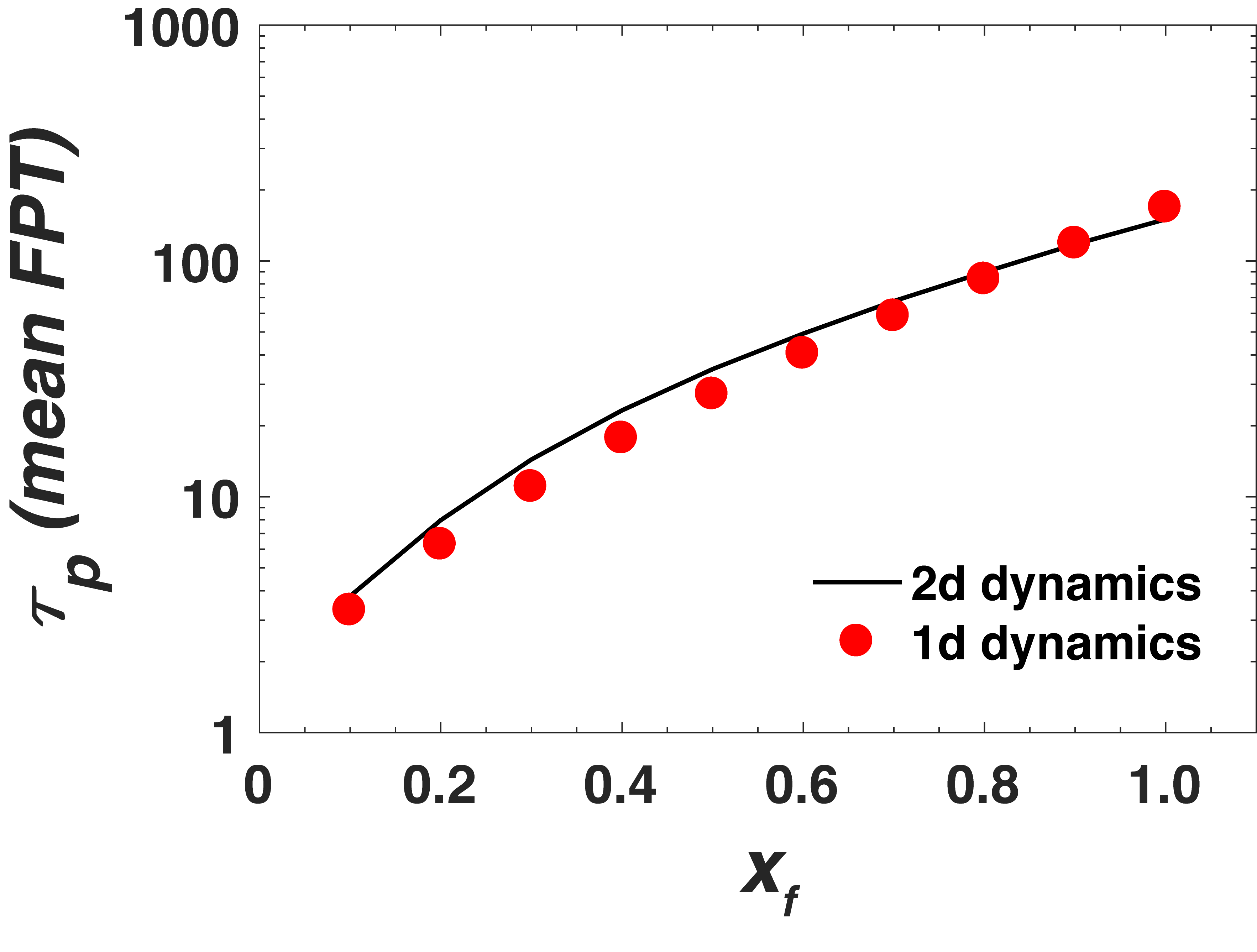}
        \caption{{\bf Position dependent diffusion coefficient captures the mean first passage time}. Comparison of the estimated mean first passage time $\tau_p$ to reach $x(t) = x_f$ for the first time when starting from $x(0) = 0$ as a function of $x_f$ in the 2 dimensional dynamics (black, Eq.~\ref{eq:lang}) and the effective one dimensional dynamics (red circles, Eq.~\ref{eq:smol1}).   \label{fg:fpt}}
\end{figure}

The effective one dimensional dynamics described by Eq.~\ref{eq:smol1} can also predict other trajectory-based dynamical quantities without any adjustible parameters. In Fig.~\ref{fg:fpt}, we show the agreement between the mean first pasage time $\tau_p$ to reach $x(t) = x_f$ for the first time when starting from $x(0) = 0$ as a function of $x_f$.

\section{Conclusion}

 We have described a method for updating a Smoluchowski equation based on observables captured in state- and path-dependent constraints.  We showed how this can be expressed in terms of position-dependent diffusion coefficients. We illustrated by considering the effective one-dimensional dynamics of a two-dimensional overdamped harmonic oscillator, with a position dependent diffusion coefficient $D(q)$.  The present approach is not limited to updating Markov models that are continuous time and continuous space; this approach can also handle discrete-time discrete-space models using Eq.~\ref{eq:k1}. Similarly, Eq.~\ref{eq:kappa} illustrates how to update a continuous time Markov process.

{\bf Acknowledgments:} KD appreciates support from the National Science Foundation (grant number 1205881) 

\providecommand*\mcitethebibliography{\thebibliography}
\csname @ifundefined\endcsname{endmcitethebibliography}
  {\let\endmcitethebibliography\endthebibliography}{}


\pagebreak
\newpage

\section{Appendix}
\setcounter{page}{1}
\setcounter{figure}{0}

\subsection{Imposing detailed balance in discrete time Markov processes\label{ap1}}
We start with Eq.~\ref{eq:caliber} in the main text. We have the Caliber,
\begin{eqnarray}
C &=&  -\sum_{a,b} y_a k_{ab} \log k_{ab} +  \sum_{a,b} y_a k_{ab} \log P_{ab} \nonumber \\&+& \sum_a l_a \left ( \sum_b y_a k_{ab} - y_a\right ) +  \sum_b m_b \left ( \sum_a y_a k_{ab} - y_b\right ) \nonumber \\
&+& \sum_{a,b} \epsilon_{ab}\left (y_a k_{ab} - y_{b}k_{ba} \right ) + \gamma \left (  \sum_{a,b} y_{a} k_{ab} r_{ab} - \bar r\right ). \label{eq:caliber1}
\end{eqnarray}
In Eq.~\ref{eq:caliber1},  Lagrange multipliers $\{ l_a \}$ impose constraints in Eq.~\ref{eq:constr1}.  $\{ m_b \}$ impose constraints in Eq.~\ref{eq:constr2}. $\{ \epsilon_{ab} \}$ impose microscopic detailed balance. Finally,  $\gamma$ imposes the constraint of ensemble average $\bar r$ of the dynamical variable $r_{ab}$ (see Eq.~\ref{eq:dyna}).
Differentiating with respect to $k_{ab}$ and setting the derivative to zero,
\begin{eqnarray}
\log \frac{k_{ab}}{P_{ab}} &=& l_a -1 + m_b + \delta_{ab} + \gamma r_{ab}
\end{eqnarray}
where $\delta_{ab} = \epsilon_{ab} - \epsilon_{ba}$. We have
\begin{eqnarray}
k_{ab} = \tau_a \lambda_b \rho_{ab} W_{ab} \label{eq:startap1}
\end{eqnarray}
where $\tau_a = \exp (l_a-1)$, $\lambda_b = \exp (m_b)$, $\rho_{ab} = \exp (\delta_{ab})$, and $W_{ab} = P_{ab}\exp \left (\gamma r_{ab} \right)$.

We impose detailed balance, $y_a k_{ab} = y_b k_{ba}$, to evalulate $\rho_{ab}$. We have
\begin{eqnarray}
y_a\tau_a \lambda_b \rho_{ab} W_{ab} &=& y_b \tau_b \lambda_a \rho_{ba} W_{ba} \\
\Rightarrow \rho_{ab}&=& \sqrt{\frac{y_b \tau_b \lambda_a P_{ba}}{y_a \tau_a \lambda_b P_{ab}}}\times  e^{\left (\gamma \frac{r_{ba} - r_{ab}}{2} \right )} \\
&=& \sqrt{\frac{y_b \tau_b \lambda_a x_a}{y_a \tau_a \lambda_b x_b}}\times  e^{\left (\gamma \frac{r_{ba} - r_{ab}}{2} \right )}  \label{eq:rhoab}
\end{eqnarray}
The last equality is a result of the fact that the Markov chain described by transition probabilities $P_{ab}$ obeys detailed balance with respect to the stationary distribution $x_a$, $x_a P_{ab} = x_b P_{ba}$. Substituting $\rho_{ab}$ in Eq.~\ref{eq:rhoab} into Eq.~\ref{eq:startap10},
\begin{eqnarray}
k_{ab} &=& \sqrt{\frac{y_b \tau_b \lambda_a x_a}{y_a \tau_a \lambda_b x_b}}\times  e^{\left (\gamma \frac{r_{ba} - r_{ab}}{2} \right )} \times \tau_a \lambda_b P_{ab} e^{\gamma r_{ab}} \\
&=& \frac{1}{y_a}  \sqrt{\tau_a\lambda_a x_a y_a} \times \sqrt{\tau_b\lambda_b x_b y_b} \frac{P_{ab}}{x_b}e^{\left (\gamma \frac{r_{ba} + r_{ab}}{2} \right )} \nonumber \\
\end{eqnarray}
We substitute $\sqrt{\tau_a\lambda_a x_a y_a} = \mu_a$ and $\frac{P_{ab}}{x_b}e^{\left (\gamma \frac{r_{ba} + r_{ab}}{2} \right )} = G_{ab}$ and we obtain Eq.~\ref{eq:k1} in the main text.

\subsection{Deriving transition rates for the continuous time Markov process\label{ap2}}

In the main text, we claimed that the transition rates for the maximum entropy continuous time Markov process with a updated equilibrium distribution $x_a \rightarrow y_a$ and after imposing additional dynamical constraints is given by Eq.~\ref{eq:kappa_a},
\begin{eqnarray}
\kappa_{ab} = \lim_{dt\rightarrow 0} \frac{k_{ab}}{dt} = \sqrt{\frac{y_b}{y_a}}\sqrt{\frac{x_a}{x_b}}\omega_{ab} \exp \left ( \gamma r_{ab} \right )\label{eq:kappa_a}.
\end{eqnarray}
Here, we prove this assertion.

Let us consider the equation
\begin{eqnarray}
D[G\bar \mu] = \bar \mu
\end{eqnarray}
where
\begin{eqnarray}
G = \frac{P_{ab}}{x_b}\exp \left ( \gamma r_{ab}\right )
\end{eqnarray}
where $r_{ab}$ is assumed to be symmetric in $a$ and $b$; $r_{ab} = r_{ba}~\forall~a~{\rm and}~b$. Plugging the transition probabilities $P_{ab}$ in Eq.~\ref{eq:pab_approx} in Eq.~\ref{eq:G} and~\ref{eq:fixd}, we have
\begin{eqnarray}
G_{ab} &=& dt \frac{\omega_{ab}}{x_b}\exp (\gamma r_{ab})~{\it if}~a\neq b\\
G_{aa} &=& \frac{1-dt \sum_{b\neq a} \omega_{ab}}{x_a} \exp (\gamma r_{aa})
\end{eqnarray}

Thus,
\begin{eqnarray}
G = J + dt\Delta
\end{eqnarray}
where $J$ is a diagonal matrix with $J_{aa} =  \exp (\gamma r_{aa})/x_a$ and
\begin{eqnarray}
\Delta_{ab} &=&  \frac{\omega_{ab}}{x_b}\exp (\gamma r_{ab})~{\it if}~a\neq b \label{eq:g1}\\
\Delta_{aa}&=&  - \frac{\sum_{b\neq a}\omega_{ab}}{x_a}\exp (\gamma r_{aa}) \label{eq:g2}
\end{eqnarray}
Substituting Eq.~\ref{eq:g1} and~\ref{eq:g2} in Eq.~\ref{eq:fixd}, we have
\begin{eqnarray}
        D[(J+dt\Delta)\bar \mu] =  \bar \mu \label{eq:fixd1}
\end{eqnarray}

Note that if $\bar \mu$ is a solution of Eq.~\ref{eq:fixd1}, we have
\begin{eqnarray}
k_{ab} &=& dt \frac{\mu_a\mu_b}{y_a}G_{ab}~{\it if}~a\neq b.
\end{eqnarray}
Thus, we need to find $\bar \mu$ only till the zeroth order in $dt$ as $dt\rightarrow 0$. We have
\begin{eqnarray}
\frac{y_a}{J_{aa} \mu_a + dt \sum_b \Delta_{ab} \mu_b} = \mu_a \label{eq:mu_a}
\end{eqnarray}
The solution for $\mu$ as $dt\rightarrow 0$ to the zeroth order of Eq.~\ref{eq:mu_a} is given by
\begin{eqnarray}
\mu_a = \sqrt{\frac{y_a}{J_{aa}}}
\end{eqnarray}
Substituting this value of $\mu_a$ in Eq.~\ref{eq:k1}, we have
\begin{eqnarray}
k_{ab} = dt\sqrt{\frac{y_b}{y_a}} \sqrt{\frac{x_a}{x_b}}\omega_{ab} \exp \left ( \gamma r^{\dag}_{ab}  \right )
\end{eqnarray}
where
\begin{eqnarray}
r^{\dag}_{ab} =   r_{ab} - \frac{r_{aa} + r_{bb}}{2}
\end{eqnarray}
is a transformed version of the dynamical constraint $r_{ab}$ such that $r_{aa}=0~\forall~a$.

\subsection{Two dimensional diffusion~\label{ap3}}

We start with the prior dynamics
\begin{eqnarray}
\frac{\partial}{\partial t} p([x,y]|t) &=& D_x\frac{\partial^2}{\partial x^2}p([x,y]|t) + D_y\frac{\partial^2}{\partial y^2}p([x,y]|t)  \nonumber \label{eq:2d_diff_ap} \\
\end{eqnarray}
In Eq.~\ref{eq:2d_diff} $D_x$ is the diffusion coefficient in the $x$ direction and $D_y$ is the diffusion coefficient in the $y$ direction.

Discretizing the space in steps of $dx=dy$ in the $x$ and $y$ direction respectively and discretizing time in steps of $dt$ (omiting the time dependence for brevity),
\begin{eqnarray}
\frac{dp([x,y])}{dt}&\approx& \omega_{([x-dx,y]),([x,y])} p([x-dx,y]) \nonumber \\ &+& \omega_{([x+dx,y]),([x,y])} p([x+dx,y]) \nonumber \\ &+& \omega_{([x,y-dy]),([x,y])} p([x,y-dy]) \nonumber \\ &+&  \omega_{([x,y+dy]),([x,y])} p([x,y+dy]) \nonumber \\
&-& p([x,y]) \omega_{[x,y],[x,y]} \label{eq:discrete_2d}
\end{eqnarray}
where
\begin{eqnarray}
\omega_{[x,y],[x,y]} &=&   \omega_{([x,y]),([x-dx,y])} + \omega_{([x,y]),([x+dx,y])} \nonumber \\ &+& \omega_{([x,y]),([x,y-dy])}  \omega_{([x,y]),([x,y+dy])}  \label{eq:self_2d}
\end{eqnarray}
In Eq.~\ref{eq:discrete_2d} $\omega_{a,b}$ denotes the transition rate of going from $a$ to $b$. We have $\omega_{a,b} = 0$ if $a$ and $b$ are not nearest neighbors on the lattice. From Eq.~\ref{eq:2d-diff}, we can write $\omega_{[x,x\pm dx,y],[x,y]} = D_x dx^2$,  $P_{[x,y\pm dy],[x,y]} = D_y dy^2$, and so on.

Next, we impose a dynamical constraint $\langle r_{ab} \rangle = \langle \phi(a) \phi(b) \rangle$ (see Eq.~\ref{eq:rmod}). Consider two points $a = [x, y]$ and $b = [w,u]$. First,  we  explicitly carry out the transformation in Eq.~\ref{eq:rdag}. We write (omitting the $\dag$ for brevity)
\begin{eqnarray}
r_{[x,y], [w, u]} &=&  \left ( \phi([x,y]) - \phi([w,u]) \right )^2 \label{eq:2dtransform}
\end{eqnarray}

From Eq.~\ref{eq:modD} and Eq.~\ref{eq:kappa_sm} We can write the updated transition rates
\begin{eqnarray}
\kappa_{[x,y],[x+dx,y]} &=& D_x dx^2 dt \exp \left  (\gamma r_{[x,y],[x+dx,y]} \right )~{\rm and} \nonumber  \\
\kappa_{[x,y],[x,y+dy]} &=& D_y dy^2 dt \exp \left  (\gamma r_{[x,y],[x,y+dy]} \right ). \label{eq:2dapprox_k}
\end{eqnarray}
Other transition probabilities can be written down similarly by recognizing that $r_{[x,y],[u,w]} = r_{[u,w],[x,y]}$. We can further simplify Eq.~\ref{eq:2dapprox_k},
\begin{eqnarray}
r_{[x,y],[x+dx,y]} &\approx& -\frac{dx^2}{2} \left ( \frac{\partial }{\partial x} \phi([x,y]) \right )^2~{\rm and} \\
r_{[x,y],[x,y+dy]} &\approx& -\frac{dy^2}{2} \left ( \frac{\partial }{\partial y} \phi([x,y]) \right )^2
\end{eqnarray}
Consequently,
\begin{eqnarray}
\kappa_{[x,y],[x+dx,y]} &=& D_x dx^2 \exp \left  (-\gamma_0 \frac{\left ( \frac{\partial }{\partial x} \phi([x,y]) \right )^2}{2} \right ), \nonumber \\
\kappa_{[x,y],[x,y+dy]} &=& D_x dx^2  \exp \left  (-\gamma_0 \frac{\left ( \frac{\partial }{\partial y} \phi([x,y]) \right )^2}{2} \right ) \nonumber \label{eq:dxdy_ap} \\
\end{eqnarray}
In Eq.~\ref{eq:dxdy_ap} have recognized $\gamma_0 = \gamma / dx^2$. Finally, the position dependent diffusion coefficients are given by
\begin{eqnarray}
D_x^{\rm new}([x,y]) = D_x\exp \left  (-\gamma_0 \frac{\left ( \frac{\partial }{\partial x} \phi([x,y]) \right )^2}{2} \right )\nonumber \\
D_y^{\rm new}([x,y]) = D_y\exp \left  (-\gamma_0 \frac{\left ( \frac{\partial }{\partial y} \phi([x,y]) \right )^2}{2} \right )
\end{eqnarray}

\subsection{Details of the Langevin dynamics~\label{ap4}}

Let us start with a Smoluchowski equation with a position dependent diffusion coefficient.
\begin{eqnarray}
\frac{\partial}{\partial t}p(x,t) = \frac{\partial}{\partial x} \left \{ D(x) e^{ -0.7x^2 } \frac{\partial}{\partial x} \left [e^{ 0.7x^2 } p(x,t) \right ]\right \} \nonumber  \label{eq:smol1a} \\
\end{eqnarray}

There are multiple ways to map this Smoluchowski equation to a Langevin equation. The two popular approaches are the Ito approach and the Stratonovich approach. Both approaches lead to the same equilibrium distribution and have the same dynamics. The  time-discretized Langevin equation with Ito convention is given by
\begin{eqnarray}
x(t+dt) &\approx& x(t) - 1.4D(x(t)) x(t) dt + \rho \sqrt{2D(x(t))dt} \nonumber \\ &+& \left (\frac{d D(x)}{dx} |_{x=x(t)} \right )dt \label{eq:1dl}
\end{eqnarray}
where $\rho$ is a normally distributed random number with mean 0 and standard deviation 1. As above, we use $dt = 0.001$ units.

\end{document}